\documentclass[12pt]{article}

\usepackage{cite}
\usepackage{amsmath,amsfonts,amssymb,amsthm,amsxtra,epsfig,epstopdf,url,array}

\usepackage{graphicx}
\usepackage{subfigure}
\usepackage{color}
\usepackage{latexsym}
\usepackage{bm}

\theoremstyle{plain}
\newtheorem{thm}{\bf Theorem}[section]
\newtheorem{lem}[thm]{\bf Lemma}

\newtheorem*{cor}{\bf Corollary}

\theoremstyle{definition}
\newtheorem{defn}{Definition}[section]

\newtheorem{exmp}{\bf Example}[section]

\theoremstyle{remark}
\newtheorem{rem}{Remark}

\def\dh#1{\mathop {#1}\limits_{h}}

\def\dh#1{ \mathop{#1}\limits_ h}

\def\dh#1{\mathop {#1}\limits_h}

\def\dvhb#1{\dh#1_{\bar 1}}
\def\dvhb2#1{\dh#1_{\bar 2}}

\def\dh#1{\mathop {#1}\limits_{h}}

\def\dvhb#1{\dh#1_{\bar x}}

\def\dh#1{\mathop {{#1}}\limits_{h}}

\def\dh#1{ \mathop{#1}\limits_ h}

\begin{document}


 \begin{center}
  {\bf \Large Lagrangian formalism and Noether-type theorems for second-order delay ODEs}
  \end{center}
\bigskip

\begin{center}
{\large Vladimir  Dorodnitsyn}$^a$,
{\large Roman Kozlov}$^b$,  \\
{\large Sergey  Meleshko}$^c$
\end{center}

\bigskip
{\bf 10.02.2023}

\vspace*{10mm}

\noindent $^a$
Keldysh Institute of Applied Mathematics, Russian Academy of Science, \\
Miusskaya Pl.~4, Moscow, 125047, Russia; \\
{e-mail:  Dorodnitsyn@Keldysh.ru, dorod2007@gmail.com} \\
$^b$ Department of Business and Management Science, Norwegian School
of Economics,
Helleveien 30, 5045, Bergen, Norway;  \\
{e-mail: Roman.Kozlov@nhh.no}  \\
$^c$ School of Mathematics, Institute of Science,
Suranaree University of Technology, 30000, Thailand; \\
{e-mail: sergey@math.sut.ac.th} \\

\begin{center}
{\bf Abstract}
\end{center}
\begin{quotation}
The Lagrangian formalism for variational problem for second-order
delay ordinary differential equations (DODEs) is developed. The
Noether-type operator identities and theorems for  DODEs of second
order are presented. Algebraic construction of integrals for
DODEs based on symmetries are demonstrated by examples.

\end{quotation}

 \noindent

\section{Introduction}

Lie group analysis is  efficient tool for studying ordinary and
partial differential equations since its introduction in the
classical work of Sophus Lie in~\cite{bk:Lie[1888],  bk:Lie1924}.
A symmetry transformation of a differential equation maps a solution
into a solution. This property is used to obtain new solutions from known
ones, and to classify equations into equivalence classes.  It can also
be used to obtain exact analytic solutions that are invariant under
some subgroup of the symmetry group (`group invariant solutions').
Applications of Lie groups to differential equations is the topic of
many books and
articles~\cite{bk:Ovsiannikov1978,bk:Olver[1986],bk:Ibragimov[1983],bk:AncoBluman1997,%
bk:HandbookLie, bk:Gaeta1994}. Since  the fundamental work of
E.Noether~\cite{Noether1918} the symmetry group becomes a starting point
to obtain first integrals and conservation laws for differential
equations, which possess Lagrangian or Hamiltonian formulation.
Recently, the connections between a symmetry group and conservation laws
for differential equations that do not have a variational formulation (and
hence they have neither a Lagrangian nor a Hamiltonian)
were developed in~\cite{bk:Ibragimov[1983],bk:BlumanAnco2002}.

More recently applications of Lie groups of transformations have
been extended to finite-difference, differential-difference,
discrete equations and integral equations~\cite{Dorodnitsyn1991, Dorodnitsyn1993,
 DorKozWin2004, bk:DKapKozWin,
bk:DKWin,LeviWinternitz2005, DorodnitsynKozlovWinternitz2000,
QuispelCapelSahadevan, bk:DKap1, bk:DKap2,  bk:Dorodnitsyn[2011],
bk:Hydon2014, bk:ChevDK, bk:DKapMel, bk:DK12, bk:DKM20, bk:DK21,bk:Meleshko[2005],bk:GrigorievIbragimovKovalevMeleshko2010}.

This article is part of a study that aims to extend group analysis
application to another type of equations: delay differential
equations. In the recent
article~\cite{bk:DorodnitsynKozlovMeleshkoWinternitz[2018a]}  there
was presented the Lie group classification of first-order delay
ordinary differential equations. Linear  first-order delay ordinary
differential equations were considered in
article~\cite{bk:DorodnitsynKozlovMeleshkoWinternitz[2018b]}. Lie
group classification of delay second-order ordinary differential
equations was studied in
~\cite{bk:DorodnitsynKozlovMeleshkoWinternitz2021}. Exact solutions
for delay PDEs were constructed in~\cite{Polian1, Polian2,Polian3}.

The purpose of this article is to construct Lagrangian formalism and
the Noether-type theorems for second-order delay ordinary
differential equations.

As shown below, the variational problem leads to an
 ODE with a delay of second-order (in the sense of the order
of a differential equation) with two delays. Thus, we have to set
an initial-value problem for delay ODEs with {\it two} delays.

We consider the formulation of the initial-value problem for a
second-order ordinary delay differential equation with two constant
delays $\tau$
\begin{equation}  \label{DODE1}
\ddot{u} =
F(\ddot{u}^{--},\ddot{u}^-,\dot{u}^-,\dot{u},\dot{u}^{--},u^-, u,
u^{--}, t,t^-,t^{--}),
 \qquad x \in I ,
\end{equation}
where $I \subset  \mathbb{R} $ is some finite or semi{finite}
interval, $t^-=t-\tau, \quad t^{--}=t- 2\tau$. Functions $u^{-}=
u(t-\tau), u^{--}= u(t-2\tau)$, while function $F$ satisfies the condition
\begin{equation}  \label{FDODE}
\left( {\partial F  \over  \partial {u}^{--} } \right) ^2 +\left(
{\partial F  \over  \partial \dot{u}^{--} } \right) ^2 +\left(
{\partial F \over  \partial \ddot{u}^{--} } \right) ^2 {\not\equiv}
0.
\end{equation}
The variable $t$ varies continuously over the entire interval, where
Eq.~(\ref{DODE1}) is defined.

In order to solve DODEs on some interval $I$ one must add initial
conditions to the DODE~(\ref{DODE1}). Contrary to the case of
ordinary differential equations, the initial condition must be given
by a smooth function $ \varphi (t)$ and it's derivative $
\dot{\varphi} (t)$ on an initial interval $ I_0 \subset \mathbb{R}
$, e.g.
\begin{equation}    \label{initialvalues}
u(t)   = \varphi (t), \qquad \dot{u}(t)   = \dot{\varphi} (t),
\qquad t \in [ t_{-2} , t_0 ] , \quad t_0- t_{-2} =2\tau, \quad t_0-
t_{-1} =\tau.
\end{equation}
For a constant $  \tau  $ this leads to the {\it method of
steps}~\cite{bk:Elsgolts[1955]} for solving the DODE either
analytically or numerically. Hence, for $ t_0 \leq t \leq t_1 = t_0 +
\tau $ we replace
$$
u^{--} = u ( t^{--} ) =   \varphi (t - 2\tau ), \quad \dot{u}^{--} =
u ( t- 2\tau) = \dot{\varphi} (t - 2\tau ), \quad \ddot{
u}^{--}=\ddot u ( t- 2\tau) = \ddot{\varphi} (t - 2\tau ),
$$
$$
u^{-} = u ( t^{-} ) =   \varphi (t - \tau ), \quad \dot{u}^{-} = u (
t- \tau) = \dot{\varphi} (t - \tau ), \quad \ddot {u}^-=\ddot u ( t-
\tau) = \ddot{\varphi} (t - \tau ),
$$
and this reduces  DODE~(\ref{DODE1}) to the ODE
\begin{equation}  \label{DODE2}
\ddot{u} = F(\ddot{\varphi} (t - 2\tau ),\dot{\varphi} (t - 2\tau ),
{\varphi} (t - 2\tau ),\ddot{\varphi} (t - \tau ),\dot{\varphi} (t -
\tau ),\varphi (t - \tau ),u,\dot u, t,t^-,t^{--}),
\end{equation}
which is solved with the initial conditions
$$
u(t_0) = \varphi (t_0  ), \quad \dot{u}(t_0) = \dot{\varphi} (t_0 ).
$$

On the second step we consider the same procedure: we solve the
DODE~(\ref{DODE1}) where $u (u - \tau ), u (u - 2\tau ) $, it's
derivatives and the initial conditions are known from the first step.
Thus, one continues 
at each step solving the ODE with data from the previous step.
 This
procedure provides a solution that is in general continuous at the
points $ t_n = t_0 + n \tau $.
\par
The method of steps for solving the initial value
problem introduces a natural sequence of intervals
\begin{equation}   \label{intervals}
[ t_{n} , t_{n+1} ] ,      \qquad      n = -1, 0, 1, 2, ...
\end{equation}
\par
 We should indicate  that peculiarity of the
initial value problem for DODEs with {\it two} delays is that the
initial conditions should be given on a double length interval
$[t_{-2} , t_0 ]$.

\bigskip
\par

This paper is devoted to analytic technique of integration of a
second-order DODE, based on its symmetry and Lagrangian formalism. The Noether-type theorems for invariant delay ordinary
differential equations  of a second-order are presented. The
integration technique based on a symmetry of an DODE is demonstrated by
examples.

The paper is organized as follows. In Section 2 we describe how to
apply symmetry generators to delay ODEs. The  analog of the first
Noether's theorem of the invariance condition for delay functional
is proved in Section 3. In Section 4 we show that the stationary
value of delay functional along a given group orbit at an arbitrary
point $t,u$ is achieved on so called local extremal equation,
special case of which is an equation named the Elsgolts equation.
The invariance of the Elsgolts equation considered in Section 5.
In section 6 we introduce the first integrals for DODE, which are sufficiently
different from those for the ODE. The main operator identities, which
allow us to formulate a Noether-type theorem for the conservation law and
symmetries are given in section 7. In section 8 we demonstrate
application of the developed methodology by examples. Short observation
of the obtained results are considered in the Conclusion.

 \noindent

 \section{Point symmetries of delay ODEs }
  \setcounter{equation}{0}

We consider a second-order delay ODE (\ref{DODE1}) and its point
symmetries defined by a generator  in the standard form:
\begin{equation} \label{operator1}
X={\xi(t,u)}{\partial {} \over \partial t}  + {\eta(t,u)} {\partial
{}   \over \partial u} + ...,
\end{equation}
which we prolong for $\dot u, \ddot u$ using standard
formulas, and for the shifted values $(t^{-},u^{-},\dot u^-),
(t^{--},u^{--},\dot u^{--}) $:
\begin{equation} \label{operator2}
\begin{array}{c}
\tilde{X}={\xi(t,u)}{\partial {} \over \partial t}  + {\eta(t,u)}
{\partial {}   \over \partial u} +{\xi^-(t^{-},u^{-})}{\partial {} \over
\partial t^{-}} + {\eta^-(t^{-},u^{-})} {\partial {}   \over \partial u^{-}}
\\[2ex]
+(D(\eta)- \dot u D(\xi))\frac{\partial {}}{\partial \dot
u}+(D^-(\eta^-)- \dot u^- D^-(\xi^-))\frac{\partial {}}{\partial
\dot u^-}+....
\\[2ex]
+({\xi(t,u)-\xi^-(t^-,u^-)}){\partial {} \over
\partial \tau}+...,
\end{array}
\end{equation}
where $\tau=t-t^-=t^{-}-t^{--}=const$, and
$$
D=\frac{\partial {}}{\partial t} +\dot{u}\frac{\partial {
}}{\partial u}  +..., \qquad D^-=\frac{\partial {}}{\partial t^-}
  +\dot{u}^- \frac{\partial { }}{\partial u^-}+....
$$
are the operators of the total derivative at points $t$ and $ t^-$,
respectively.

 We also use the operators
\[
D^{+}=\frac{\partial}{\partial
t^{+}}+\dot{u}^{+}\frac{\partial}{\partial
u^{+}}+\ddot{u}^{+}\frac{\partial}{\partial\dot{u}^{+}}+...,
\]
and the following sum of operators
\[
\bar{D}=D^{-}+D+D^{+}.
\]

We split the operator of the total derivative into three
different operators for preserving their point character.
In this case they can be multiplied (from the left) by any functions $\xi(t,u)$,
$\xi^-(t^-,u^-)$ and $\xi^+(t^+,u^+)$, respectively, without loss
the nature of its tangent conditions (for belonging to a higher order
group or Lie-B\"acklund group, see \cite{Dorodnitsyn1993a}).
\par
\bigskip

We also use a pair of shift operators to the right and left hand-sides,
respectively:
\begin{equation}
S_+ (f(t,u))= f(t+\tau, u^+)=f^+,  \qquad S^{-} (f(t, u))= f(t-\tau,
u^-)=f^-.
\end{equation}
The shift operators can be expressed using the operators of the
total derivative $D$:
$$
S_+={\sum^{}_{s\geq0}{\frac{\tau ^s}{s!}}D^s}, \quad
S^{-}={\sum^{}_{s\geq0}{\frac{(-\tau) ^s}{s!}}(D^-)^s}.
$$

Here and below, it is assumed that the delay parameter $\tau=const$. To keep
regular intervals invariant with respect to generator
(\ref{operator1}) we should consider those of them, which satisfy the
condition (see \cite{bk:Dorodnitsyn[2011]}):
\begin{equation} \label{Regular}
(S_+ -1)(S_{-} -1) [\xi(t,u)]=0,
\end{equation}
or
\begin{equation} \label{Regular2}
\xi^+ -2\xi +\xi^-=0.
\end{equation}
Thus, the criterion of the invariance of equation (\ref{DODE1})
reads:
\begin{equation} \label{crit}
\tilde{X}(\ddot{u} -
F(\ddot{u}^{-},\dot{u}^{-},\dot{u},u,u^{-},t,...))_{|_{(\ref{DODE1})}}=0,
\end{equation}
provided by condition (\ref{Regular2}).

\bigskip

\section{The invariance of delay Lagrangians }
  \setcounter{equation}{0}

Let a one-parameter group of point transformations be:
\begin{equation}\label{Group}
t_* = f(t,u,a), \quad u_*=g(t,u,a),
\quad{t_*}^-=f^-(t^-,u^-,a),\quad {u_*}^-=g^-(t^-,u^-,a),
\end{equation}
which are prolonged for derivatives with the help of standard
formulas of Lie group analysis
\cite{bk:Ovsiannikov1978,bk:Olver[1986],bk:Ibragimov[1983]}. We need
also a group action on the differential:
\begin{equation}
dt_* = Df(t,u,a)da.
\end{equation}
Considering delay functional
\begin{equation}
\label{eq:feb09.1}
L= \int_{a}^{b}{\cal L}(t,u,u^-,\dot{u},\dot{u}^-)\, dt,
\end{equation}
we chose any inner interval $a+\tau <t_1< t_2<b $ for the anaysis:
\begin{equation}
L= \int_{t_1}^{t_2}{\cal L}(t,u,u^-,\dot{u},\dot{u}^-)\, dt.
\end{equation}

\par
 The functional is referred as invariant, if
\begin{equation}
L= \int_{t_1}^{t_2}{\cal L}(t,u,u^-,\dot{u},\dot{u}^-)\, dt
=\int_{t^*_1}^{t^*_2}{\cal
L}(t_*,u_*,{u_*}^-,{\dot{u}}_*,{\dot{u}_*}^-)\, dt^*.
\end{equation}
The transformed variables and the interval of integration are represented by
means of the initial values and the group parameter
\begin{equation}
\int_{t_1}^{t_2}{\cal L}(t,u,u^-,\dot{u},\dot{u}^-)\, dt
=\int_{t_1}^{t_2}{\cal L}(f,g,g^-,\dot{g},\dot{g}^-,
)\,J(\frac{f}{t}) dt,
\end{equation}
where $J(\frac{f}{t})$ is the Jacobian of the transformation
(\ref{Group}).

As the interval is arbitrary one can omit the integration:
\begin{equation}\label{Group1}
{\cal L}(t,u,u^-,\dot{u},\dot{u}^-) dt ={\cal
L}(f,g,g^-,\dot{g},\dot{g}^-, ) J(\frac{f}{t})dt.
\end{equation}

Hence, the variational integral is invariant if and only if the
elementary action is invariant with respect to  (\ref{Group1}).

Differentiating (\ref{Group1}) with respect to a group parameter $a$,
and setting $a=0$, we obtain a criterion for a delay functional to be
invariant.

\par
\begin{thm}\label{thm_D1}
 (First Noether's theorem for delay ODEs)
\par
\bigskip
Functional (\ref{eq:feb09.1}) is invariant if and only if
\begin{equation}\label{Group2}
\xi\frac{\partial {\cal L}}{\partial t} + \xi^-\frac{\partial {\cal
L}}{\partial {t^{-}}} + \eta\frac{\partial {\cal L}}{\partial u} +
\eta^-\frac{\partial {\cal  L}}{\partial  {u^-}} +
\end{equation}
$$
+(D(\eta)- \dot u D(\xi))\frac{\partial {\cal L}}{\partial \dot
u}+(D^-(\eta^-)- \dot u^- D^-(\xi^-))\frac{\partial {\cal
L}}{\partial \dot u^-}+ {\cal L} D(\xi)=0.
$$
\end{thm}

We should note that, in contrast to the classical case of ODEs, there are
{\it two} group orbits passing through  the elementary action ${\cal
L}(t,u,u^-,\dot{u},\dot{u}^-) dt$. When $\tau \rightarrow 0$ the criterion (\ref{Group2}) becomes the classical first Noether's theorem.

\par
\bigskip
\begin{rem}
Paper~\cite{art:FredericoTorres} (see also \cite{Second
Noether'stheorem}) contains wrong formula for the invariance of a
delay Lagrangian. Therefore, the subsequent formulas and
formulations are also incorrect.
\end{rem}

 \section{The Elsgolts extremal equation and local extremal equation}
\par
\bigskip
\par
Consider the delay functional (\ref{eq:feb09.1})

\begin{equation}
L= \int_{a}^{b}{\cal L}(t,t^-,u,u^-,\dot{u},\dot{u}^-)\, dt
\end{equation}
and apply the variational operator along a fixed curve (for
considerations of directional variation see
\cite{bk:RozhdYanenko[1978]}).

\par

The variation of all variables is applied {\it only} along  the
group orbit at the internal point $(t,u), a<t<b$:
$$
\delta t = \xi \delta a,\quad \delta u = \eta \delta a, \quad
\delta\dot{u} = ({D}(\eta)- \dot u {D}(\xi)) \delta a,\quad \delta
(dt)= {D}(\xi)dt\delta a,
$$
where $\delta a$ is the variation of the group parameter $a$.

\par
 Variation of the functional only along a group orbit yields:
\begin{equation}
\begin{array}{c}
 \delta L = \int_{a}^{b}\left[\left(\frac{\partial {\cal L}}{\partial t}\delta t+
\frac{\partial {\cal L}^+}{\partial {t}} \delta t +
  \frac{\partial {\cal L}}{\partial u}
 \delta u+\frac{\partial {\cal L^+}}{\partial u} \delta u
+ \frac{\partial {\cal L}}{\partial \dot{u}}
 \delta \dot{u} + \frac{\partial {\cal L^+}}{\partial \dot{u}}
 \delta \dot{u}\right)dt +  {\cal L}\delta (dt)\right]=
\\[2ex]
=\int_{a}^{b}\left(\xi \left(\frac{\partial {\cal L}}{\partial t}+
\frac{\partial {\cal L}^+}{\partial {t}}\right) +
\eta\left(\frac{\partial {\cal L}}{\partial u} +\frac{\partial {\cal
L^+}}{\partial u}\right) +({D}(\eta)- \dot u {D}(\xi))
\left(\frac{\partial {\cal L}}{\partial \dot{u}}
 + \frac{\partial {\cal L^+}}{\partial \dot{u}}\right)+{\cal L}{D}(\xi)\right)
dt\delta a.
\end{array}
\end{equation}
Integration by parts gives (here boundary values are not varied):
\begin{equation}
\label{eq:feb09.2}
\begin{array}{c}
 \delta L =\int_{a}^{b}\xi \left( { \partial {\cal L} \over \partial t} + { \partial
{\cal L} ^+ \over
\partial t } + D  \left( \dot{u}  { \partial {\cal L} \over \partial \dot{u}
} \right) + D  \left( \dot{u}  { \partial {\cal L} ^+ \over
\partial \dot{u} } \right) - D ({\cal L})\right)+
\\[2ex]
+\eta \left(
 { \partial {\cal L} \over \partial u}
 +  { \partial {\cal L}^+  \over \partial u }
- D  \left(  { \partial {\cal L} \over \partial \dot{u} }
\right) - D  \left( { \partial {\cal L} ^+ \over \partial  {
\dot{u}  } } \right)
 \right)dt\delta a.
\end{array}
\end{equation}

\par
The integral is equal to zero for arbitrary interval of integration if
the integrand is zero. Hence, we arrive at {\it the
local extremal equation}:
\begin{equation}\label{localextremal}
\begin{array}{c}
\xi \left( { \partial {\cal L} \over \partial t} + { \partial {\cal
L} ^+ \over
\partial t } + D  \left( \dot{u}  { \partial {\cal L} \over \partial \dot{u}
} \right) + D  \left( \dot{u}  { \partial {\cal L} ^+ \over
\partial \dot{u} } \right) - D ({\cal L})\right)
\\[2pt]
+\eta \left(
 { \partial {\cal L} \over \partial u}
 +  { \partial {\cal L}^+  \over \partial u }
- D  \left(  { \partial {\cal L} \over \partial \dot{u} }
\right) - D  \left( { \partial {\cal L} ^+ \over \partial  {
\dot{u}  } } \right)
 \right) = 0,
 \end{array}
\end{equation}
which {\it explicitly depends} on $\xi, \eta$, i.e. on a given
group.
\bigskip
\par

\par
\bigskip
\par
Notice, that a local extremal equation contains $\cal L $ and $ \cal
L^+$, and therefore, the equation contains {\it two delays}. The
operator ${D}$ is applied at the point $(t,u)$, and the
directional variation only occurs along the group orbit passing through
the point $(t,u)$.
\par

\bigskip

\par

"Vertical variation", i.e. variation by $u$ when $\xi=0$,  yields
the following  extremal delay equation
\begin{equation}\label{Euler}
\frac{\partial {\cal L}}{\partial u}- {D} (\frac{\partial {\cal
L}}{\partial \dot{u}}) +\frac{\partial {\cal L^+}}{\partial u} -
{D}( \frac{\partial {\cal L^+}}{\partial \dot{u}})=\frac{\delta
}{\delta u}({\cal L}   +{\cal L^+}) =0,
\end{equation}
 This
equation is known since Elsgolts  \cite{bk:Elsgolts[1955]} (see also
 \cite{Elsgolts2}, \cite{Elsgolts3} ).
 We will call it {\it the Elsgolts equation}.

 We call {\it the Elsgolts variational
derivative} the following operator
\begin{equation}\label{Elsgolts}
{\frac{\delta }{\delta u}}_{(E)}     =\frac{\partial {}}{\partial
u}-{D} \frac{\partial {}}{\partial \dot{u}} + S_+ \left(
\frac{\partial {}}{\partial u^-}-{D} \frac{\partial {}}{\partial
\dot{u}^-} \right),
\end{equation}
which one can apply to a delay Lagrangian ${\cal L}$. Notice, that the
shift operator $S_+$ acts on {\it all} arguments.

The Elsgolts equation operates with ${\cal L}$ and ${\cal L}^+$,
while the criterion of invariance of a Lagrangian only involves
$\cal L$. We should stress  that in contrast to a case of ODEs and
PDEs,  the Elsgolts equation is generally not an
equation, on a solution of which the Lagrangian achieves its
extremal value.

We also separate out "horizontal" variation
\begin{equation}\label{Horizont}
{\frac{\delta }{\delta t}} ({\cal L}   +{\cal L^+}) = {
\partial {\cal L} \over
\partial t} + {
\partial {\cal L} ^+ \over
\partial t } + {D} \left( \dot{u}  { \partial {\cal L} \over \partial \dot{u}
} \right) + {D} \left( \dot{u}  { \partial {\cal L} ^+ \over
\partial \dot{u} } \right) - {D}({\cal L})=0,
 \end{equation}
which yields a variational equation for the case $\eta$=0.

The horizontal variational operator is:
\begin{equation}   \label{operator_variational_t}
{ \delta  \over \delta t} = { \partial    \over \partial t} + {D}
\left( \dot{u}  { \partial       \over \partial \dot{u} } \right) +
S _+ \left( { \partial       \over \partial t ^-} + {D} \left(
\dot{u} { \partial       \over \partial \dot{u} ^- } \right) \right)
- {D}.
\end{equation}
Now one can rewrite the local extremal equation as
\begin{equation}         \label{quasi}
\xi   { \delta L \over \delta t } +  \eta { \delta L \over \delta u
}_{(E)} = 0.
\end{equation}

The local extremal equation gives a necessary condition for a
Lagrangian to achieve extremal value on the direction of a given
group. The invariance of a Lagrangian is not needed (for directional
variation see \cite{bk:RozhdYanenko[1978]}). Connection of the
invariance of a Lagrangian with local extremal equation is the subject of the next section.
\bigskip

\par
\par

\section{Invariance of the Elsgolts equation}

Consider the generator
\[
X=\xi(t,u)\partial_{t}+\eta(t,u)\partial_{u}
\]
preserving constant delay.

\begin{lem} Let the coefficient $\xi(t)$ of the generator
\[
X=\xi(t)\partial_{t}+\eta(t,u)\partial_{u}
\]
satisfy condition (\ref{Regular2}) and
${\dot{\xi}_t}={\dot{\xi^+}_{t^+}} $. The following identity holds:
\begin{equation}
\frac{\delta}{\delta u}_{(E)}\left(X{\cal L}+{\cal
L}{D}(\xi)\right)=X\left(\frac{\delta{\cal L}}{\delta
u}_{(E)}\right)+(\eta_{u}+{D}(\xi)-\dot{u}\xi_{u})\frac{\delta{\cal
L}}{\delta u}_{(E)}.\label{lemma2}
\end{equation}
\end{lem}

The proof can be obtained by direct calculations.

In more details relation (\ref{lemma2}) can be rewritten as
\begin{equation}
\begin{array}{c}
\left(\frac{\partial{}}{\partial u}-\bar{D}\frac{\partial{}}{\partial\dot{u}}\right)\left(X{\cal L}+{\cal L}\bar{D}(\xi)
+X{\cal L}^{+}+{\cal L}^{+}\bar{D}(\xi^{+})\right)
\\[2ex]
=X\left(\left(\frac{\partial{}}{\partial u}-\bar{D}\frac{\partial{}}{\partial\dot{u}}\right)\left({\cal L}+
{\cal L}^{+}\right)\right)+\left(\left(\frac{\partial{}}{\partial u}-\bar{D}\frac{\partial{}}{\partial\dot{u}}\right)
\left({\cal L}+{\cal L}^{+}\right)\right)({\eta_{u}}+\bar{D}(\xi)+\dot{u}\xi_{u}).
\end{array}\label{lemma2-1}
\end{equation}

From identity (\ref{lemma2}) one can establish the relation between
the invariance of a Lagrangian and the invariance of the corresponding
Elsgolts equation.

  Considering Eq.(\ref{lemma2}) on a solution of the Elsgolts equation
 one has
\[
{\frac{\delta}{\delta u}}_{(E)}\left(X{\cal L}+{\cal
L}{D}(\xi)\right)|_{\frac{\delta{\cal L}}{\delta
u}_{(E)}=0}=X\left(\frac{\delta{\cal L}}{\delta
u}_{(E)}\right)|_{\frac{\delta{\cal L}}{\delta u}_{(E)}=0}=0.
\]

\bigskip{}

\begin{thm} \label{th:1} Let the generator $X$ satisfy property
(\ref{Regular2}) and ${\dot{\xi}_t}={\dot{\xi^+}_{t^+}} $. Then the
Elsgolts equation is invariant if and only if the following
condition holds:
\begin{equation}\label{theorem12}
\frac{\delta}{\delta u}_{(E)}\left(X{\cal L}+{\cal
L}{D}(\xi)\right)|_{\frac{\delta{\cal L}}{\delta u}_{(E)}=0}=0
\end{equation} for any solution of the Elsgolts equation. \end{thm}

Note that Theorem \ref{th:1} well matches with the analogous theorem for
ODE \cite{bk:DorodnitsynKozlov[2011]}.
\bigskip{}

\begin{exmp} Consider the Lie algebra $L_{4.13}$ \cite{bk:GonzalezKamranOlver[1992b]}
of the generators
\begin{equation}
X_{1}=\frac{\partial}{\partial
u},\,\,\,X_{2}=\frac{\partial}{\partial
t},\,\,\,X_{3}=t\frac{\partial} {\partial
t},\,\,\,X_{4}=u\frac{\partial}{\partial u},\label{eq:ex1.may29}
\end{equation}
which preserves the uniformity of delay parameter $\tau$.

 The complete set
of invariants of this Lie algebra consists of the invariants
\[
I_{1}=\frac{u-u^{-}}{\dot{u}(t-t^{-})},\,\,\,I_{2}=\frac{\dot{u}^{-}}{\dot{u}},\,\,\,\,I_{3}=(t-t^{-})\frac{\ddot{u}}{\dot{u}}.
\]
As all generators (\ref{eq:ex1.may29}) satisfy property
(\ref{theorem12}), the Elsgolts equation
 with the Lagrangian ${\cal L}={\cal
L}(I_{1},I_{2})$ admits the generators (\ref{eq:ex1.may29}).
Meanwhile the Lagrangian  ${\cal L}={\cal L}(I_{1},I_{2})$ does not
admit the scaling generator $X_{3}=t\frac{\partial}{\partial t}$.
\end{exmp}

\section{First integrals for DODEs}

\bigskip
\par
The standard form of a solution (integral) of an ODE
\begin{equation}\label{ODE}
I(t,u)= A,
\end{equation}
can not satisfy any initial data which are given on an initial
interval for DODEs. Hence, we have to introduce another form of an
integral.
\par

DODEs  (\ref{DODE1}) can contain the dependent variable and its
derivatives at three points $t^+ $, $ t$  and $t^-$. It is possible
to consider two types of conserved quantities: differential first
integrals and difference first integrals.

\begin{defn}   \label{differential_first_integral}
Quantity
\begin{equation}
I  (   t  ^+,   t, t ^- ,
 u  ^+  ,  u,  u  ^-  ,
 \dot{u}  ^+  ,   \dot{u}  , \dot{u}  ^-     )= A_0
\label{1}
\end{equation}
is called {\it a differential first integral} of DODE (\ref{DODE1})
if it holds constant on the solutions of the DODE.
\end{defn}

Consideration of differential first integrals needs differentiation
and  requires to chose constant delays. We have to define
differentiation of functions, which depend on the dependent variables for
several values of the independent variable. Differentiating $I$ at
some point, for example, at the point $t$, we get
\begin{equation*}
\bar{D}  (I) =
  ( I _{t ^+ }  + I _{u  ^+ }  \dot{u} ^+  + I _{  \dot{u} ^+ }
\ddot{u} ^+ )   { d t  ^+  \over d t } + ( I _t + I _u   \dot{u}  +
I _{  \dot{u} }    \ddot{u}  ) +   ( I _{t ^- }  + I _{u  ^- }
\dot{u} ^-  + I _{  \dot{u} ^- }
  \ddot{u} ^- )   { d t ^-  \over d t },
\end{equation*}
where
\begin{multline}    \label{differentiation}
\bar{D} =
 {\partial \over \partial t}
+  \dot{u}   {\partial  \over \partial u} +  \ddot{u}   {\partial
\over \partial \dot{u}} + \cdots + {\partial \over \partial t^- } +
\dot{u}   ^-  {\partial  \over \partial u^- } +   \ddot{u}  ^-
{\partial  \over \partial \dot{u}^- } + \cdots
\\
+  {\partial \over \partial t^+ } +   \dot{u}  ^+  {\partial  \over
\partial u^+ } +   \ddot{u}  ^+   {\partial  \over \partial
\dot{u}^+ } + \cdots
\end{multline}

Generally, the delay can depend on the independent variable $t$ as
well as on the solution $ u(t) $ of the DODE. Then, the values $ t^+ $
and $ t^- $ have the same dependence. It makes sense to restrict the
considered delays by the conditions
\begin{equation}  \label{delay_conditions}
{  d t ^+ \over d t }  = 1 , \qquad {  d t ^-  \over d t }  = 1
\end{equation}
for all considered $t$. In this case the differentiation does not
depend on a chosen point, and  we obtain that  the differential
first integrals should satisfy the equation
\begin{equation}
\bar{D} ( I ) =   I _{t ^+ }    +  I _t      +   I _{t ^- } + I _{u
^+ } \dot{u} ^+    + I _u   \dot{u}    + I _{u  ^- }  \dot{u} ^- + I
_{ \dot{u} ^+ }    \ddot{u} ^+    + I _{  \dot{u} }    \ddot{u}   +
I _{  \dot{u} ^- }    \ddot{u} ^- = 0 ,
\end{equation}
which should hold on the solutions of the considered DODE.

Conditions    (\ref{delay_conditions}) imply that the delays
should be solution independent. Moreover, integrating these
conditions,  we obtain the delay equations   that justifies the
choice of the constant delays for Eq. (\ref{DODE1}). If we tried to
consider {non}constant delays, we would have difficulties with
defining differentiation and differential first integrals.

\begin{defn}   \label{deference_first_integral}
A function
\begin{equation}
J  (     t, t ^- ,
      u,  u  ^-  ,
     \dot{u}  , \dot{u}  ^- ,
\ddot{u}  , \ddot{u}  ^-       )
\end{equation}
is called {\it a difference first integral} of DODE (\ref{DODE1}) if
it satisfies the equation
\begin{equation}
( S_+   - 1 )   J  = 0
\end{equation}
on solutions of the DODE.
\end{defn}

We illustrate  the definitions of the first integrals on a simple
example.

\begin{exmp}
DODE
\begin{equation*}
  \ddot{u} ^+ =     \ddot{u} ^-
\end{equation*}
has differential first integral
\begin{equation*}
 I =  \dot{u} ^+ -   \dot{u} ^-,
\end{equation*}
and difference first integral
\begin{equation*}
  J  =  \ddot{u}   -    \ddot{u} ^-    .
\end{equation*}
\end{exmp}

Differential  first integrals are constant on the solutions of
DODEs. In this case, the difference first integral should not be
constant on DODE solutions, they can be periodic
functions with period $\tau$, where $\tau$ is the delay parameter.

\par
\bigskip
 \noindent
\section{The  Noether-type identity for delay Lagrangians}

\par
\bigskip
In this Section we introduce the main operator identity, which relates
together the invariance of a Lagrangian and first integral of local extremal equation.
\par
   We consider a Lagrangian function
\begin{equation}
{\cal L} = {\cal L}(t,t^-,u,u^-,\dot{u},\dot{u}^-),
\end{equation}
which depends on $t, t^-$, and we associate its invariance with
first integral.

\par
\bigskip
\begin{lem}\label{lem_D3}
The following identity holds ($\tau$=const):

\begin{equation}\label{identity1}
\begin{array}{c}
\xi\frac{\partial {\cal L}}{\partial t} +\xi^-\frac{\partial {\cal
L}}{\partial t^-} +\eta\frac{\partial {\cal L}}{\partial u} +
\eta^-\frac{\partial {\cal  L}}{\partial {u^-}} +({D}(\eta)- \dot u
{D}(\xi))\frac{\partial {\cal L}}{\partial \dot u}+
\\[2ex]
+(D^-(\eta^-)- \dot u^- D^-(\xi^-))\frac{\partial {\cal L}}{\partial
\dot u^-}+ {\cal L} {D}(\xi)\equiv
\\[2ex]
\equiv \eta \left(
 { \partial {\cal L} \over \partial u}
 +  { \partial {\cal L}^+  \over \partial u }
- {D} \left(  { \partial {\cal L} \over \partial \dot{u} } \right) -
{D} \left( { \partial {\cal L} ^+ \over \partial  { \dot{u}  } }
\right) \right)+
\\[2ex]
+ \xi \left( { \partial {\cal L} \over \partial t} + { \partial {\cal L} ^+ \over
\partial t } + {D} \left( \dot{u}  { \partial {\cal L} \over \partial \dot{u}
} \right) + {D} \left( \dot{u}  { \partial {\cal L} ^+ \over
\partial \dot{u} } \right) - {D}({\cal L})
 \right)+
\\[2ex]
+  \bar{D} \left[ \xi {\cal L}
 + ( \eta - \dot{u} \xi )   \left( { \partial {\cal L} \over \partial \dot{u} }
  +   { \partial {\cal L} ^+  \over \partial \dot{u} } \right)
\right]+
\\[2ex]
+( 1- S_+)\left( \xi^-  \frac{\partial {\cal L}}{\partial t^-} +
\eta^-  \frac{\partial {\cal L}}{\partial u^-}+({D}^-(\eta^-) -
\dot{u}^- {D}^-(\xi^-))
  \frac{\partial {\cal L}}{\partial \dot{u}^-}
\right)=0.
\end{array}
\end{equation}

\end{lem}

\par
\bigskip
Identity (\ref{identity1}) allows one to specify
different DODEs on the right-hand side and formulate various appropriate
theorems (a similar idea was implemented for ODEs in
\cite{{bk:DorodnitsynIbragimov}}).

\par
\bigskip
\par
\begin{thm}\label{thm_D3}
  ({\bf Delay version of Noether's theorem for
a local extremal  equation})

Let a functional ${\cal L}$ be invariant with respect to a
one-parameter group of transformations. Then the invariance of the Lagrangian gives the integral
\begin{equation}\label{differC}
  \bar{D} \left[ \xi {\cal L}
 + ( \eta - \dot{u} \xi )   \left( { \partial {\cal L} \over \partial \dot{u} }
  +   { \partial {\cal L} ^+  \over \partial \dot{u} } \right)
\right]=0
\end{equation}
for each solution of a local extremal  equation (\ref{quasi}),
provided by the difference relation
\begin{equation}\label{differInt}
  \xi^-  \frac{\partial {\cal L}}{\partial t^-} +
\eta^-  \frac{\partial {\cal L}}{\partial u^-}+({D}^-(\eta^-) -
\dot{u}^- {D}^-(\xi^-))
  \frac{\partial {\cal L}}{\partial \dot{u}^-}=
\end{equation}
$$
=\xi  \frac{\partial {\cal L^+}}{\partial t} + \eta \frac{\partial
{\cal L^+}}{\partial u}+({D}(\eta) - \dot{u} {D}(\xi))
  \frac{\partial {\cal L^+}}{\partial \dot{u}}.
$$

\end{thm}

\par
The difference integral (or periodic condition) relates
corresponding values at two points and is not a constant for each
point of the considered interval.

\par
\bigskip
\begin{rem}
It should be noted that the local extremal equation (\ref{quasi})
depends on second-order derivatives, and usually contains {\it
two} delays.
\end{rem}

\bigskip
\par
{\bf Consequences   of Lemma \ref{lem_D3} and Theorem \ref{thm_D3}}
\par
\bigskip
The \ref{thm_D3} theorem has an obvious generalization for the following
cases:

\par
\bigskip
1. If an action of a symmetry operator on an elementary Lagrangian
action ${\cal L}dt$ gives $ \bar{D}(A(t,t^-,u,u^-,\dot u,\dot
u^-))$, then one can add the corresponding function $(- A)$  into the
expression of local extremal equation. This is a generalization
similar to that for ODE \cite{Bessel_Hagen}.
\par
\bigskip

2. If an action of a symmetry operator on an elementary Lagrangian
action ${\cal L}dt$ gives $ ( S_+ -1)(B(t,t^-,u,u^-,\dot u,\dot
u^-))$, then one can add the corresponding function $(- B)$  into the
expression of difference integral (periodic condition).
\par
Both cases can exist in one action of a symmetry operator on an
elementary Lagrangian action ${\cal L}dt$.
\par

\bigskip
3. It may happens that expression (\ref{differInt}) is equivalent to
some divergent term $ \bar{D}(C(t,t^-,u,u^-,\dot u,\dot u^-))$:
\begin{equation}\label{dif21}
  \xi^-  \frac{\partial {\cal L}}{\partial t^-} +
\eta^-  \frac{\partial {\cal L}}{\partial u^-}+({D}^-(\eta^-) -
\dot{u}^- {D}^-(\xi^-))
  \frac{\partial {\cal L}}{\partial \dot{u}^-}-
\end{equation}
$$
-\xi  \frac{\partial {\cal L^+}}{\partial t} - \eta \frac{\partial
{\cal L^+}}{\partial u}-({D}(\eta) - \dot{u} {D}(\xi))
  \frac{\partial {\cal L^+}}{\partial \dot{u}} =\bar{D}(C),
$$
 then
we arrive to the case 1 (see examples of the linear oscillator
below).
\par
\bigskip
Below we consider two special cases of Noether's theorem for two
special cases of admitted group, which are in the examples.

\par
\begin{cor}
 ({\bf First special version of Noether's theorem for
delay ODEs})

Let a functional ${\cal L}$ be invariant with respect to a
one-parameter group of transformations with $\xi =0$. Then the
invariance of the Lagrangian gives the integral
\begin{equation}\label{a1}
\bar{D} \left[
 \eta    \left( { \partial {\cal L} \over \partial \dot{u} }
  +   { \partial {\cal L} ^+  \over \partial \dot{u} } \right)
\right]=0
\end{equation}
for each solution of the Elsgolts equation
\begin{equation}\label{a2}
 { \partial {\cal L} \over \partial u}
 +  { \partial {\cal L}^+  \over \partial u }
- \bar{D} \left(  { \partial {\cal L} \over \partial \dot{u} }
\right) - \bar{D} \left( { \partial {\cal L} ^+ \over \partial  {
\dot{u}  } } \right) =0,
\end{equation}
under the following periodic condition (difference integral)
\begin{equation}\label{differInt3}
 \eta^-  \frac{\partial {\cal L}}{\partial u^-}+({D}^-(\eta^-) -
\dot{u}^- {D}^-(\xi^-))
  \frac{\partial {\cal L}}{\partial \dot{u}^-}=
\end{equation}
$$
= \eta \frac{\partial {\cal L^+}}{\partial u}+({D}(\eta) - \dot{u}
{D}(\xi))
  \frac{\partial {\cal L^+}}{\partial \dot{u}}.
$$

\end{cor}

\par
\begin{cor}
 ({\bf Second special version of Noether's theorem for
delay ODEs})

Let a functional ${\cal L}$ be invariant with respect to a
one-parameter group of transformations with $\eta =0$. Then the
invariance of Lagrangian yields the integral
\begin{equation}\label{b1}
\bar{D} \left[ \xi {\cal L}
 +   \dot{u} \xi    \left( { \partial {\cal L} \over \partial \dot{u} }
  +   { \partial {\cal L} ^+  \over \partial \dot{u} } \right)
\right]=0
\end{equation}
for each solution of the following equation
\begin{equation}\label{b2}
 { \partial {\cal L} \over \partial t} + { \partial {\cal
L} ^+ \over
\partial t } + \bar{D} \left( \dot{u}  { \partial {\cal L} \over \partial \dot{u}
} \right) + \bar{D} \left( \dot{u}  { \partial {\cal L} ^+ \over
\partial \dot{u} } \right) - \bar{D}({\cal L})
  =0,
\end{equation}
under the following periodic condition (difference integral)
\begin{equation}\label{differInt5}
\xi^-  \frac{\partial {\cal L}}{\partial t^-} - \dot{u}^-
{D}^-(\xi^-)
  \frac{\partial {\cal L}}{\partial \dot{u}^-}=\xi  \frac{\partial {\cal L^+}}{\partial t} - \dot{u}
{D}(\xi)
  \frac{\partial {\cal L^+}}{\partial \dot{u}}.
\end{equation}

\end{cor}
\par
\bigskip

\begin{rem}

It should be noted that  the Theorems formulated above
yield restriction (periodic conditions) on the set of solutions of
the corresponding  equations (\ref{a2}) and (\ref{b2}).

\end{rem}
\par
\bigskip

\begin{rem}
 It may happens that expression (\ref{differInt3}) or (\ref{differInt5}) is
equivalent to some divergent term $ \bar{D}(C)$, then we add the
term $-\bar{D}(C)$ to integrals (\ref{a1}), (\ref{b1}). In these
cases we have no periodic restrictions for the set of solutions.
\end{rem}

\vfil
\eject
\par
\section{Examples of applications}
\par

\par
{ \bf 1. Linear oscillator 1}

Consider the following delay Lagrangian
\begin{equation}
{\cal L}={u}{u}^{-}-\dot{u}\dot{u}^{-},
\label{Lag1}
\end{equation}
which is divergently invariant with respect to the generator  $X_{1}={\cos t\frac{\partial}{\partial u}}$:
\[
X_{1}({\cal L})=u^{-}\cos t+u\cos t^{-}+\dot{u}^{-}\sin t+\dot{u}\sin t^{-}=\bar{D}[u^{-}\sin t+u\sin t^{-}].
\]

 Notice that this symmetry allows one to consider the
constant $\tau$ for the whole interval.

According to Theorem \ref{thm_D3}, we get the following difference
integral
\begin{equation}
u\cos t^{-}+\dot{u}\sin t^{-}=u^{+}\cos t+\dot{u}^{+}\sin t.\label{E1}
\end{equation}
Taking into account divergence term in the action of $X_{1}$ on the
Lagrangian, we get the following differential first integral
\begin{equation}
\bar{D}(\cos t(\dot{u}^{+}+\dot{u}^{-})+u\sin t^{-}+{u}^{-}\sin t)=0,\label{inte1}
\end{equation}
for solutions of the Elsgolts equation
\begin{equation}
{\frac{\delta{\cal L}}{\delta
u}}_{(E)}=\ddot{u}^{+}+\ddot{u}^{-}+u^{+}+u^{-}=0.\label{Elsg1}
\end{equation}
The first integral (\ref{inte1}) is equivalent to the Elsgolts equation
(\ref{Elsg1}) by virtue  of difference relation (\ref{E1}). It is
easy to check that eq. (\ref{Elsg1}) is invariant with respect to
$X_{1}$.

\bigskip{}

The delay Lagrangian (\ref{Lag1}) is divergently invariant
with respect to the generator  $X_{2}={\sin t\frac{\partial}{\partial u}}$:
\[
X_{2}({\cal L})=u^{-}\sin t+u\sin t^{-}-\dot{u}^{-}\cos t-\dot{u}\cos t^{-}=-\bar{D}[u^{-}\cos t+u\cos t^{-}].
\]

According to Theorem \ref{thm_D3}, we derive the difference integral
\begin{equation}
u\sin t^{-}-\dot{u}\cos t^{-}=u^{+}\sin t-\dot{u}^{+}\cos
t.\label{E21}
\end{equation}
Taking into account the divergent term in the action of $X_{2}$ on the Lagrangian
elementary action, we get the following differential first integral
\begin{equation}
\bar{D}\left(-\sin t(\dot{u}^{+}+\dot{u}^{-})+u\cos t^{-}+{u}^{-}\cos t\right)=0,\label{inte2}
\end{equation}
for a solution of the same Elsgolts eq. (\ref{Elsg1}). Notice that
eq. (\ref{Elsg1}) is invariant with respect to $X_{2}$.

The first integral (\ref{inte2}) is equivalent to  Elsgolts eq.
(\ref{Elsg1}) by virtue  of difference relation (\ref{E21}).

\bigskip{}

Thus, for Elsgolts eq. (\ref{Elsg1}) we derived two differential
delay integrals
\begin{equation}
\cos t(\dot{u}^{+}+\dot{u}^{-})+u\sin t^{-}+{u}^{-}\sin t=A,\label{integrals}
\end{equation}
\[
-\sin t(\dot{u}^{+}+\dot{u}^{-})+u\cos t^{-}+{u}^{-}\cos t=B,
\]
from which one can obtain the integral of the form (\ref{1})
\begin{equation}
u\cos\tau+{u}^{-}=A\sin t+B\cos t.\label{const}
\end{equation}
Notice that we have used both differential and difference integrals.

\begin{rem}

Consider an oscillating solution of the Elsgolts eq. (\ref{Els1})
of the form
\begin{equation}
u=\alpha\sin t+\beta\cos t,\quad\alpha,\beta=const,\label{sol1}
\end{equation}
that means that we also take
\begin{equation}
u^{-}=\alpha\sin(t-\tau)+\beta\cos(t-\tau)\label{sol2}
\end{equation}
as the initial data together with (\ref{sol1}). Hence, $u$ and $u^{-}$
are given as initial conditions on the intervals $[-2\tau,-\tau]$
and $[-\tau,0]$ respectively, while $u^{+}$ is a solution to be
found.

One can show that in this case the solution
\begin{equation}
u^{+}=\alpha\sin(t+\tau)+\beta\cos(t+\tau)\label{sol3}
\end{equation}
solves the Elsgolts eq. (\ref{Elsg1}). Substituting the solutions
(\ref{sol1}), (\ref{sol2}) into integral (\ref{const}), one obtains the
relations for the constants:
\begin{equation}
2\alpha\cos\tau+\beta\sin\tau=A,\label{relait}
\end{equation}
\[
2\beta\cos\tau+\alpha\sin\tau=B.
\]

The Elsgolts eq. (\ref{Elsg1}) also possesses
infinitely many solutions besides special oscillating solution
(\ref{sol1}).

\end{rem}
\bigskip{}

\par
{ \bf 2. An alternative approach to linear oscillator}
\bigskip{}

Consider the same Lagrangian $ {\cal
L}={u}{u}^{-}-\dot{u}\dot{u}^{-}$, which is divergently
invariant with respect to the generator $X_{1}={\cos
t\frac{\partial}{\partial u} + \cos t^-\frac{\partial}{\partial u^-}
- \sin t\frac{\partial}{\partial \dot{u}} - \sin
t^-\frac{\partial}{\partial \dot{u^-}}}$.

The difference expression
\begin{equation}
u\cos t^{-}+\dot{u}\sin t^{-} - u^{+}\cos t - \dot{u}^{+}\sin t =
\bar{D}({u}\sin t^{-} - {u}^{+}\sin t)
\end{equation}
is equivalent to exact divergence of some expression for the underlined
case.  We add this divergent term into the differential first integral
\begin{equation}\label{intel1}
-\bar{D}(\cos t(\dot{u}^{+}+\dot{u}^{-}) + \sin t(u^- + u^+)) =
\end{equation}
$$
-\cos t(\ddot{u}^{+}+\ddot{u}^{-}+u^{+}+u^{-})=0
$$
 for the Elsgolts equation
\begin{equation}
{\frac{\delta{\cal L}}{\delta
u}}_{(E)}=\ddot{u}^{+}+\ddot{u}^{-}+u^{+}+u^{-}=0.\label{Els1}
\end{equation}

\bigskip{}

The delay Lagrangian (\ref{Lag1}) is also divergently invariant
with respect to the generator  $X_{2}={\sin
t\frac{\partial}{\partial u}}$:
\[
X_{2}({\cal L})=u^{-}\sin t+u\sin t^{-}-\dot{u}^{-}\cos
t-\dot{u}\cos t^{-}=-\bar{D}[u^{-}\cos t+u\cos t^{-}].
\]

The difference expression is
\begin{equation}
u\sin t^{-}-\dot{u}\cos t^{-} - u^{+}\sin t + \dot{u}^{+}\cos
t=\bar{D} (u^+ \cos t - u \cos t^-)  \label{E2}
\end{equation}
Taking into account the latter divergent term, we get the
following differential first integral:
\begin{equation}
\bar{D}\left(-\sin t(\dot{u}^{+}+\dot{u}^{-}) +\cos t(u^- + u^+)
\right)=  -\sin t
(\ddot{u}^{+}+\ddot{u}^{-}+u^{+}+u^{-})=0,\label{inte21}
\end{equation}
which is equivalent to  Elsgolts eq. (\ref{Els1}).
\bigskip{}

Thus, for Elsgolts eq. (\ref{Els1}) we derived two differential
 integrals
\[
\begin{array}{c}
\cos t(\dot{u}^{+}+\dot{u}^{-}) + \sin t(u^- +
u^+)=A,
\\[2ex]
-\sin t(\dot{u}^{+}+\dot{u}^{-}) +\cos t(u^- + u^+) =B,
\end{array}
\]
from which one can obtain the integral of the form (\ref{1})
\begin{equation}
u^+ +{u}^{-}=A\sin t+B\cos t.\label{const1}
\end{equation}
Notice that we have no restrictions for integral (\ref{const1}).

\par
\bigskip
{ \bf 3. Nonlinear oscillator 1 }

\par
\bigskip

 Lagrangian (\ref{Lag1}) is also invariant with respect to
$X_{3}={\frac{\partial}{\partial t}}$ and for the local extremal
equation
\begin{equation}
\ddot{u}\dot{u}^{+}+\ddot{u}^{+}\dot{u}+\dot{u}u^{-}+\dot{u}^{-}u=0
\label{horizont1}
\end{equation}
we get zero difference integral and the following differential one
\[
\bar{D}[\dot{u}\dot{u}^{+}+uu^{-}]=0.
\]
Thus, the symmetry $X_{3}$ yields another {\it nonlinear} local
extremal equation and corresponding differential integral without
periodic restriction for the set of solutions of Eq. (\ref{horizont1}).

\bigskip{}

{ \bf 4.  Nonlinear oscillator 2}

\bigskip{}

Consider the  delay Lagrangian
\[
{\cal L}={u}{u}^{-}-\dot{u}\dot{u}^{-},
\]
which is divergently invariant with respect to the sum of the generators
$X_{1}+ X_{3}={\frac{\partial}{\partial t}}+{\cos
t\frac{\partial}{\partial u}}$:
\begin{equation}
(X_{1}+ X_{3}){\cal L} = u^{-}\cos t+u\cos t^{-}+\dot{u}^{-}\sin
t+\dot{u}\sin t^{-}=\bar{D}[u^{-}\sin t+u\sin t^{-}].
\end{equation}
In this case, the local extremal equation takes the form  of the {\it first-order}
DODE:
\begin{equation}
\sin t(\dot{u}^{+}+\dot{u}^{-})+ \cos t ({u}^{-} + {u}^{+}) =0.
\end{equation}
According to Theorem \ref{thm_D3}, we get the following difference
integral
\begin{equation}
u\cos t^{-}+\dot{u}\sin t^{-} + u^{+}\cos t-\dot{u}^{+}\sin t=0.
\end{equation}
Taking into account divergent term in action of $X_{1}+X_{3}$ on
the Lagrangian, we can rewrite the local extremal equation in the
form
\begin{equation}
{D}({u}^{+} \sin t - u \sin t^{-})=0,
\end{equation}
which is equivalent to the difference integral.

Thus, a solution can be found from the relation
\begin{equation}
 {u}^{+} \sin t - u \sin t^{-} = A_0
\end{equation}
and initial data.

\bigskip{}

{ \textbf{5. Linear oscillator 2} }

\bigskip{}

Consider the following delay Lagrangian
\begin{equation}
{\cal L}=\frac{(\dot{u}+\dot{u}^{-})^{2}}{2}-\frac{(u+u^{-})^{2}}{2},\label{Lag2}
\end{equation}
which is divergently invariant with respect to the generator  $X_{1}={\cos t\frac{\partial}{\partial u}}$:
\[
X_{1}({\cal L})=-(u+u^{-})(\cos t+\cos t^{-})-(\dot{u}^{-}+\dot{u})(\sin t+\sin t^{-})=-\bar{D}[(u+u^{-})(\sin t+\sin t^{-})].
\]

According to Theorem \ref{thm_D3}, we obtain the following difference integral
\begin{equation}
(u+u^{-})\cos t^{-}+(\dot{u}+\dot{u}^{-})\sin t^{-}=(u+u^{+})\cos t+(\dot{u}+\dot{u}^{+})\sin t.\label{E21a}
\end{equation}
Taking into account the divergence term in the action of $X_{1}$ on the Lagrangian,
we get the following differential first integral
\begin{equation}
\bar{D}[\cos t(\dot{u}^{+}+2\dot{u}+\dot{u}^{-})+(u+{u}^{-})(\sin t^{-}+\sin t)]=0,\label{inte2}
\end{equation}
for a solution of the Elsgolts equation
\begin{equation}
{\frac{\delta{\cal L}}{\delta u}}_{(E)}=\ddot{u}^{+}+2\ddot{u}+\ddot{u}^{-}+u^{+}+2u+u^{-}=0.
\label{Els21b}
\end{equation}
The first integral (\ref{inte2}) is equivalent to  the Elsgolts eq. (\ref{Els21b})
by virtue  of difference relation (\ref{E21a}). One can check that
Eq. (\ref{Els21b}) is invariant with respect to $X_{1}$.

\bigskip{}

The same delay Lagrangian is divergently invariant
with respect to the generator  $X_{2}={\sin t\frac{\partial}{\partial u}}$:
\[
X_{2}({\cal L})=(\dot{u}^{-}+\dot{u})(\cos t+\cos t^{-})-(u+u^{-})(\sin t+\sin t^{-})=\bar{D}[u+{u}^{-})(\cos t+\cos t^{-})].
\]

Theorem \ref{thm_D3} yields the following difference integral
\begin{equation}
-\sin t^{-}({u}^{-}+{u})+\cos t^{-}(\dot{u}+\dot{u}^{-})=-\sin t({u}^{+}+{u})+\cos t(\dot{u}+\dot{u}^{+}).\label{d2}
\end{equation}
Taking into account the divergent term in the action of $X_{2}$ on the Lagrangian
elementary action, we get the following differential first integral
\begin{equation}
\bar{D}[\sin t(\dot{u}^{+}+2\dot{u}+\dot{u}^{-})-(u+u^{-})(\cos t^{-}+\cos t)]=0,\label{integ2}
\end{equation}
for a solution of the same Elsgolts Eq. (\ref{Els21b}). Eq. (\ref{Els21b}) is also invariant with respect to $X_{2}$.

The first integral (\ref{integ2}) is equivalent to the  Elsgolts Eq. (\ref{Els21b})
by virtue  of the difference relation (\ref{d2}).

\bigskip{}

Thus, for the Elsgolts Eq. (\ref{Els21b}) we have got two differential
delay integrals
\begin{equation}
\cos t(\dot{u}^{+}+2\dot{u}+\dot{u}^{-})+(u+{u}^{-})(\sin t^{-}+\sin t)=A,\label{integral1}
\end{equation}
\begin{equation}
\sin t(\dot{u}^{+}+2\dot{u}+\dot{u}^{-})-(u+u^{-})(\cos t^{-}+\cos t)=B,\label{integral2}
\end{equation}
from which one can obtain the integral of the form
\begin{equation}
(u+{u}^{-})(1+\cos\tau)=A\sin t-B\cos t.\label{solution2}
\end{equation}
Again, we have used both differential and difference integrals.

\begin{rem}
There exists a special oscillating solution of the Elsgolts
eq. (\ref{Els21b})
\begin{equation}
u=\alpha\sin t+\beta\cos t,\quad\alpha,\beta=const,\label{sol11}
\end{equation}

with the corresponding initial values and the relations for the
constants:
\begin{equation}
(\alpha+\alpha\cos\tau+\beta\sin\tau)(1+\cos\tau)=A,
\label{relait2}
\end{equation}
\[
(\beta+\beta\cos\tau-\alpha\sin\tau)(1+\cos\tau)=-B.
\]
\end{rem}

{ \textbf{ 6. Nonlinear DODEs} }

\bigskip{}

Consider the following delay Lagrangian
\begin{equation}
{\cal L}=\frac{({u}-{u}^{-})^{2}}{\dot{u}\dot{u}^{-}},
\label{Lag3}
\end{equation}
which is invariant with respect to the following generators:
\begin{equation}
X_{1}={\frac{\partial}{\partial u}};\quad X_{2}={u\frac{\partial}{\partial u}};\quad X_{3}={u^{2}\frac{\partial}{\partial u}};\quad X_{3}={\frac{\partial}{\partial t}}.\label{oper3}
\end{equation}

1. Using the generator $X_{1}$, and according to Theorem \ref{thm_D3}, we
have the following difference relation:
\begin{equation}
\frac{{u}-{u}^{-}}{\dot{u}\dot{u}^{-}}=\frac{{u}^{+}-{u}}{\dot{u}\dot{u}^{+}},
\label{dif3}
\end{equation}
and the following differential first integral
\begin{equation}
\bar{D}\left(\frac{({u}-{u}^{-})^{2}}{\dot{u}^{2}\dot{u}^{-}}+\frac{({u}^{+}-{u})^{2}}{\dot{u}^{2}\dot{u}^{+}}\right)=0,\label{inte3}
\end{equation}
for a solution of the Elsgolts equation
\begin{equation}
{\frac{\delta{\cal L}}{\delta u}}_{(E)}=\frac{2({u}-{u}^{-})}{\dot{u}\dot{u}^{-}}-\frac{2({u}^{+}-{u})}{\dot{u}\dot{u}^{+}}
+\bar{D}\left(\frac{({u}-{u}^{-})^{2}}{\dot{u}^{2}\dot{u}^{-}}+\frac{({u}^{+}
-{u})^{2}}{\dot{u}^{2}\dot{u}^{+}}\right)=0.\label{Els3}
\end{equation}
The first integral (\ref{inte3}) is equivalent to  Elsgolts eq. (\ref{Els3})
by virtue  of difference relation (\ref{dif3}), which provides restrictions for a set of solutions of  Elsgolts eq. (\ref{Els3}).

One can check that eq. (\ref{Els3}) is invariant with respect to $X_{1}$.

\bigskip{}

\bigskip{}

2. The delay Lagrangian (\ref{Lag3}) admits the generator $X_{2}={u\frac{\partial}{\partial u}}$.

According to Theorem \ref{thm_D3}, we derive the difference relation:
\begin{equation}
u^{-}\frac{2({u}-{u}^{-})}{\dot{u}^{2}\dot{u}^{-}}
+\frac{({u}-{u}^{-})^{2}}{\dot{u}\dot{u}^{-}}=u\frac{2({u}^{+}
-{u})}{\dot{u}^{2}\dot{u}^{+}}+\frac{({u}^{+}-{u})^{2}}{\dot{u}\dot{u}^{+}},\label{dif4}
\end{equation}
and the following differential first integral
\begin{equation}\label{inte4}
\bar{D}\left(\frac{({u}-{u}^{-})^{2}}{\dot{u}^{2}\dot{u}^{-}}+\frac{({u}^{+}
-{u})^{2}}{\dot{u}^{2}\dot{u}^{+}}\right)=0,
\end{equation}
for solutions of the Elsgolts equation (\ref{Els3}).

By virtue  of difference relation (\ref{dif4}), the first integral (\ref{inte4}) is equivalent to the Elsgolts eq. (\ref{Els3}).

One can check that eq. (\ref{Els3}) is invariant with respect to $X_{2}$.

\bigskip{}

3. The delay Lagrangian (\ref{Lag3}) also admits the generator $X_{3}={u^{2}\frac{\partial}{\partial u}}$, which gives the difference relation:
\begin{equation}
(u^{-})^{2}\frac{({u}-{u}^{-})}{\dot{u}^{-}}
+u^{-}\frac{({u}-{u}^{-})}{\dot{u}^{-}}=u^{2}\frac{({u}^{+}
-{u})}{\dot{u}}+u\frac{({u}^{+}-{u})}{\dot{u}},\label{dif5}
\end{equation}
and the following differential first integral
\begin{equation}
\bar{D}\left(u^{2}\left(\frac{({u}-{u}^{-})^{2}}{\dot{u}^{2}\dot{u}^{-}}+\frac{({u}^{+}-{u})^{2}}{\dot{u}^{2}\dot{u}^{+}}\right)\right)=0,\label{inte5}
\end{equation}
for a solution of the Elsgolts equation (\ref{Els3}) multiplied by $u^{2}$.
The first integral (\ref{inte5}) is equivalent to  Elsgolts eq. (\ref{Els3})
by virtue  of difference relation (\ref{dif5}), which gives constrains for a set of solutions of the Elsgolts equation.

One can check that eq. (\ref{Els3}) is invariant with respect to $X_{3}$.

\bigskip{}

\bigskip{}

4. The Lagrangian (\ref{Lag3}) is also invariant with respect to
$X_{4}={\frac{\partial}{\partial t}}$. In this case we have another
local extremal equation
\begin{equation}
\bar{D}\left(\frac{2({u}-{u}^{-})^{2}}{\dot{u}\dot{u}^{-}}+\frac{({u}^{+}-{u})}{\dot{u}\dot{u}^{+}}\right)=0,\label{horizont4}
\end{equation}
which also admits $X_{4}$. In this case difference relation vanishes.
The following differential integral holds
\begin{equation}
\frac{2({u}-{u}^{-})^{2}}{\dot{u}\dot{u}^{-}}+\frac{({u}^{+}-{u})}{\dot{u}\dot{u}^{+}}=const.\label{diffInt}
\end{equation}
Thus, the symmetry $X_{4}$ yields another nonlinear local extremal
equation and corresponding differential integral, which is valid for
all solutions of eq. (\ref{horizont4}), \bigskip{}

\bigskip{}

\section{{\large{}Conclusion}}

Lagrangian formalism for variational problem of second-order
delay differential equations (DODEs) was considered in the present
paper. The invariance condition for delay functional was established
and the delay analogue of first Noether's theorem was proved. It was
shown that the variation of a functional along the only group orbit
gives some equation (named a local extremal equation), which
explicitly depends on a group. For `vertical variation' a local extremal equation degenerates to the Elsgolts equation named after mathematician, who was the first to
derive it. For `horizontal' variation one gets some other DODEs. The
necessary and sufficient condition for the Elsgolts equation to be
invariant is established. The local extremal equation and the
Elsgolts equation have in general {\it two} delay parameters and it
is far from common consideration of DODEs in the literature. The
appropriate initial value problem for a DODE with two delays was
formulated. The generalization of the term a `first integral' for
DODEs has been given.

The basic Noether-type operator identity and delay analogues of the
Noether theorem for invariant of second-order delay ordinary differential
equations  are presented. The integration
technique of DODEs based on symmetry is demonstrated by examples.

\subsection*{Dedication}

The article is dedicated to the memory of Pavel Winternitz in
recognition of his diverse contributions to applications of Lie
group methods. The authors were happy to collaborate with him for
many years.


\begin{thebibliography}{10}

\bibitem{bk:Lie[1888]}
S.~Lie.
\newblock {K}lassifikation und {I}ntegration von gew\"ohnlichen
  {D}ifferentialgleichungen zwischen $x,y$, die eine {G}ruppe von
  {T}ransformationen gestatten. {I, II}.
\newblock {\em Math. Ann.}, 32:213--281, 1888.
\newblock Gesammelte Abhandlungen, vol. 5, B.G. Teubner, Leipzig, 1924, pp.
  240--310.

\bibitem{bk:Lie1924}
S.~Lie.
\newblock Gruppenregister.
\newblock {\em Gesammelte Abhandlungen}, 5:767--773, 1924.

\bibitem{bk:Ovsiannikov1978}
L.~V. Ovsiannikov.
\newblock {\em Group Analysis of Differential Equations}.
\newblock Nauka, Moscow, 1978.
\newblock {E}nglish translation, {A}mes, {W}.{F}., Ed., published by Academic
  Press, New York, 1982.

\bibitem{bk:Olver[1986]}
P.~J. Olver.
\newblock {\em Applications of {L}ie Groups to Differential Equations}.
\newblock Springer-Verlag, New York, 1986.

\bibitem{bk:Ibragimov[1983]}
N.~H. Ibragimov.
\newblock {\em Transformation Groups Applied to Mathematical Physics}.
\newblock Nauka, Moscow, 1983.
\newblock {E}nglish translation, Reidel, D., Ed., Dordrecht, 1985.

\bibitem{bk:AncoBluman1997}
S.~C. Anco and G.~Bluman.
\newblock Direct construction of conservation laws from field equations.
\newblock {\em Phys. Rev. Lett.}, 78(3):2869--2873, 1997.

\bibitem{bk:HandbookLie}
N.~H. Ibragimov, editor.
\newblock {\em {CRC} Handbook of {L}ie Group Analysis of Differential
  Equations}, volume 1, 2, 3.
\newblock CRC Press, Boca Raton, 1994, 1995, 1996.

\bibitem{bk:Gaeta1994}
G.~Gaeta.
\newblock {\em Nonlinear Symmetries and Nonlinear Equations}.
\newblock Kluwer, Dordrecht, 1994.

\bibitem{Noether1918}
E.~Noether.
\newblock Invariante variations problem.
\newblock {\em Nachr. d. {K}\"oniglichen Gesellschaft der Wissenschaften zu
  G\"ottingen, Nachrichten, Mathematisch-Physikalische Klasse Heft 2}, pages
  235--257, 1918.
\newblock English translation : Transport Theory and Statist. Phys., 1(3),
  1971, 183-207, (arXiv:physics/0503066 [physics.hist-ph]).

\bibitem{bk:BlumanAnco2002}
G.~W. Bluman and S.~C. Anco.
\newblock {\em Symmetry and Integration Methods for Differential Equations}.
\newblock Springer, New York, 2002.

\bibitem{Dorodnitsyn1991}
V.~A. Dorodnitsyn.
\newblock Transformation groups in net spaces.
\newblock {\em Journal of Soviet Mathematics}, 55(1):1490--1517, 1991.

\bibitem{Dorodnitsyn1993}
V.~A. Dorodnitsyn.
\newblock Finite-difference models entirely inheriting symmetry of original
  differential equations.
\newblock In {\em Modern Group Analysis: Advanced Analytical and Computational
  Methods in Mathematical Physics}, volume 191. Kluwer Academic Publishers,
  Boston, 1993.

\bibitem{DorKozWin2004}
V.~Dorodnitsyn, R.~Kozlov, and P.~Winternitz.
\newblock Continuous symmetries of lagrangians and exact solutions of discrete
  equations.
\newblock {\em J. Math. Phys.}, 45(1):336--359, 2004.



\bibitem{LeviWinternitz2005}
Decio Levi and Pavel Winternitz.
\newblock Continuous symmetries of difference equations.
\newblock {\em Journal of Physics A: Mathematical and General}, 39(2):R1--R63,
  dec 2005.

\bibitem{DorodnitsynKozlovWinternitz2000}
V.~Dorodnitsyn, R.~Kozlov, and P.~Winternitz.
\newblock Lie group classification of second-order ordinary difference
  equations.
\newblock {\em J. Math. Phys.}, 41(1):480--504, 2000.

\bibitem{QuispelCapelSahadevan}
G.~R.~W. Quispel, H.~W. Capel, and R.~Sahadevan.
\newblock Continuous symmetries of differential--difference equations.
\newblock {\em Phys. Lett.}, 170A:379--383, 1992.

\bibitem{bk:Dorodnitsyn[2011]}
V.~A. Dorodnitsyn.
\newblock {\em Applications of Lie Groups to Difference Equations}.
\newblock CRC Press, Boca Raton, 2011.

\bibitem{bk:Hydon2014}
P.~E. Hydon.
\newblock {\em Difference Equations by Differential Equation Methods}.
\newblock Cambridge University Press, Cambridge, 2014.



\bibitem{bk:DKapKozWin}
V.A.Dorodnitsyn, E.I.Kaptsov, R.V.Kozlov, and P.Winternitz. The
adjoint equation method for constructing first integrals of
difference
 equations.
 Journal of Physics A: Mathematical and Theoretical, 48(5):055202, 01 2015.


\bibitem{bk:DKap1}
V. A. Dorodnitsyn and E. I. Kaptsov. Shallow water equations in
Lagrangian coordinates: Symmetries, conservation laws and its
preservation in difference models. Communications in Nonlinear
Science and Numerical Simulation, 89:105343, 2020.


\bibitem{bk:DKap2}
V. A. Dorodnitsyn and E. I. Kaptsov. Discrete shallow water
equations preserving symmetries and conservation laws. Journal of
Mathematical Physics, 2021, { \it J. Math. Phys. }


\bibitem{bk:DKapMel}
V. A. Dorodnitsyn, E. I. Kaptsov, and S. V. Meleshko. Symmetries,
conservation laws, invariant solutions and difference schemes of the
one-dimensional Green-Naghdi equations. Journal of Nonlinear
Mathematical Physics,  Vol. 28(1);  (2021), pp. 90-107.

\bibitem{bk:DKWin}
V. A. Dorodnitsyn, R. V. Kozlov, and P. Winternitz. Continuous
symmetries of Lagrangians and exact solutions of discrete equations.
Journal of Mathematical Physics, 45(1):336-359, 2004.


\bibitem{bk:ChevDK}
A.F.Cheviakov, V.A.Dorodnitsyn, E.I.Kaptsov. Invariant conservation
law-preserving discretizations of linear and nonlinear wave
equations. Journal of Mathematical Physics, 61(8):081504, 2020.




\bibitem{bk:DK12}
 Dorodnitsyn V.A. and Kaptsov E.I.,  "Invariant Finite-Difference
Schemes for Plane One-Dimensional MHD Flows That Preserve
Conservation Laws" Mathematics, MDPI, 2022.


\bibitem{bk:DKM20}
 Dorodnitsyn V., Kaptsov E., Meleshko S., Conservative Invariant Finite-Difference
 Schemes for the Modified
Shallow Water Equations in Lagrangian Coordinates,  Studies in
Applied Mathematics, 2022.


\bibitem{bk:DK21}
 Dorodnitsyn V. A., Kaptsov E. I., Invariant finite-difference
schemes for plane one-dimensional MHD flows that preserve
conservation laws. Mathematics, 2022.

\bibitem{bk:Meleshko[2005]}
S.~V. Meleshko.
\newblock {\em Methods for Constructing Exact Solutions of Partial Differential
  Equations}.
\newblock Mathematical and Analytical Techniques with Applications to
  Engineering. Springer, New York, 2005.

\bibitem{bk:GrigorievIbragimovKovalevMeleshko2010}
Yu.~N. Grigoriev, N.~H. Ibragimov, V.~F. Kovalev, and S.~V. Meleshko.
\newblock {\em Symmetries of integro-differential equations and their
  applications in mechanics and plasma physics}.
\newblock Lecture Notes in Physics, Vol. 806. Springer, Berlin / Heidelberg,
  2010.

\bibitem{bk:DorodnitsynKozlovMeleshkoWinternitz[2018a]}
V.~A. Dorodnitsyn, R.~Kozlov, S.~V. Meleshko, and P.~Winternitz.
\newblock Lie group classification of first-order delay ordinary differential
  equations.
\newblock {\em Journal of Physics A: Mathematical and Theoretical}, 51(205202),
  2018.

\bibitem{bk:DorodnitsynKozlovMeleshkoWinternitz[2018b]}
V.~A. Dorodnitsyn, R.~Kozlov, S.~V. Meleshko, and P.~Winternitz.
\newblock Linear or linearizable first-order delay ordinary differential
  equations and their lie point symmetries.
\newblock {\em Journal of Physics A: Mathematical and Theoretical}, 51(205203),
  2018.

\bibitem{bk:DorodnitsynKozlovMeleshkoWinternitz2021}
V.~A. Dorodnitsyn, R.~Kozlov, S.~V. Meleshko, and P.~Winternitz.
\newblock Second-order delay ordinary differential equations, their symmetries
  and application to a traffic problem.
\newblock {\em Journal of Physics A: Mathematical and Theoretical}.
\newblock  IOP Publishing Ltd Volume 54, Number 10, 2021.


\bibitem{Polian1}
Polyanin A.D., Zhurov A.I., Exact separable solutions of delay
reaction-diffusion equations and other nonlinear partial
functional-differential equations, Commun. Nonlinear Sci. Numer.
Simul., 2014, Vol. 19, 409-416.


\bibitem{Polian2}
Polyanin A.D., Zhurov A.I., Nonlinear delay reaction-diffusion
equations with varying transfer coefficients: Exact methods and new
solutions, Appl. Math. Lett., 2014, Vol. 37, 43-48.


\bibitem{Polian3}
Polyanin A.D., Sorokin V.G., Reductions and exact solutions of
nonlinear wave-type PDEs with proportional and more complex delays,
Mathematics, 2023, Vol. 11, 516.



\bibitem{bk:Elsgolts[1955]}
L.~E. Elsgolts.
\newblock {\em Qualative Methods in Mathematical Analysis}.
\newblock GITTL, Moscow, 1955.
\newblock Translaition: American Mathematical Society, Providence ,1964.

\bibitem{Dorodnitsyn1993a}
V.~A. Dorodnitsyn.
\newblock The finite-difference analogy of Noether's theorem.
\newblock {\em Doklady RAN}, 328(6):678--682, 1993.
\newblock Translation in English in {\it Phys. Dokl.} {\bf 38} (2) 66--68.

\bibitem{art:FredericoTorres}
S.~F. Frederico and F.~M. Torres.
\newblock Noether's symmetry theorem for variational and optimal control
  problems with time delay.
\newblock {\em Numerical Algebra, Control and Optimization (NACO)},
  2(3):619--630, 2012.


\bibitem{Second Noether'stheorem}
 Agnieszka B. Malinowska and Tatiana Odzijewicz (2016): Second Noether's theorem with time delay,
 Applicable Analysis, DOI: 10.1080/00036811.2016.1192136


\bibitem{bk:RozhdYanenko[1978]}
B.~L. Rozhdestvenskii and N.~N. Yanenko.
\newblock {\em Systems of quasilinear equations and their applications to gas
  dynamics, 2nd ed.}
\newblock Nauka, Moscow, 1978.
\newblock {E}nglish translation published by Amer. Math. Soc., Providence, RI,
  1983.

\bibitem{bk:DorodnitsynKozlov[2011]}
V.~A. Dorodnitsyn and R.~Kozlov.
\newblock {L}agrangian and {H}amiltonian formalism for discrete equations:
  Symmetries and first integrals.
\newblock In {\em Symmetries and Integrability of Difference Equations}, pages
  7--49. Cambridge University Press, Cambridge, 2011.
\newblock London Mathematical Society Lecture Notes.



\bibitem{bk:DorodnitsynIbragimov}
V. Dorodnitsyn and N.Ibragimov. An extension of the Noether theorem:
accompanying equations possessing conservation laws.  Communications
in Nonlinear Science and Numerical Simulation, 2014, v.19, N2,
328-336.

\bibitem{bk:GonzalezKamranOlver[1992b]}
A.~Gonzalez-Lopez, N.~Kamran, and P.~J. Olver.
\newblock Lie algebras of vector fields in the real plane.
\newblock {\em Proc. London Math. Soc.}, 64:339--368, 1992.

\bibitem{Bessel_Hagen}
E.~Bessel-Hagen.
\newblock \"{U}ber die {E}rhaltungssatze der {E}lektrodynamik.
\newblock {\em Math. Ann.}, 84:258--276, 1921.

\bibitem{Elsgolts2}
L.~E. Elsgolts.
\newblock Variational problems with retarded arguments.
\newblock {\em Vestnik Moskovskogo Universiteta. Seriya 1. Matematika.
  Mekhanika}, 10:57--62, 1952.
\newblock (in Russian).

\bibitem{Elsgolts3}
L.~E. Elsgolts.
\newblock Variational problems with retarded arguments.
\newblock {\em Uspekhi Matematicheskikh Nauk}, 12(1(73)):257--258, 1957.
\newblock (in Russian).

\end{thebibliography}

\end{document}